\documentclass[9pt]{article}

\RequirePackage{amssymb}[1995/01/01]

\title{\bf{The unity between quantum field computation, real computation,
and quantum computation}}

\author{A.~C.~Manoharan \footnote{e-mail: Manoharan@worldnet.att.net}}

\begin{document}

\maketitle

California State University Stanislaus,
 Turlock, CA 95382 USA

\begin{abstract}
\bf{
It is indicated that principal models of computation
are indeed significantly related.  The quantum field computation model
contains the quantum computation model of Feynman.  
  (The term "quantum field
computer" was used by Freedman.)  
Quantum field computation (as enhanced by Wightman's model of
quantum field theory) involves computation over the continuum which is
remarkably related to the real computation model of Smale.  The latter model
was established as a generalization of Turing computation.  All this is not
surprising since it is well known that the physics of quantum field theory
(which includes Einstein's special relativity) contains quantum mechanics
which in turn contains classical mechanics.  The unity of these computing
models, which seem to have grown largely independently, could shed new light
into questions of computational complexity, 
into the central P
(Polynomial time) versus NP (Non-deterministic Polynomial time) problem of
computer science, and also into the description of Nature by fundamental
physics theories.}

\end{abstract}

\bibliographystyle{unsrt} 

\section{Introduction} 

The two great physical theories of the twentieth century were {\it quantum
theory} and {\it relativity}, both of which generalize classical Newtonian
mechanics.  Quantum mechanics and classical mechanics are special limiting
cases of quantum field theory.
By taking the limit as the
velocity of light $c \rightarrow {\infty}$ we expect to get non-relativistic
quantum mechanics.  The limit as Planck's constant $h \rightarrow 0$ gives
classical mechanics~\cite{Jaffe}. 
  Neither quantum theory nor relativity can be ignored.  Quantum field theory
is a logical and natural result of combining quantum theory and relativity.

In the same twentieth century, the mathematical theory of {\it
computation} was developed and the {\it electronic computer} was invented. 
Gate implementation for the standard classical computer or {\it Turing machine}
can be based on classical mechanics; in the sense that the gates, ideally in
the absence of perturbations (depicted by the electrical engineering term {\it
noise}), could even consist of perfect billiard balls.

Meanwhile, we have become increasingly dependent on computing machines, and
there are more models of computation.  By Church's thesis, to state it simply,
all {\em reasonable} models of computation are equivalent. 
   At the present
time, there appear to be three  principal models of computation, which have
grown largely independently.  The models are, {\it quantum field computation},
{\it real computation} and {\it quantum computation}.  It is proposed and
argued here that these three models of computation are indeed
significantly related.  Hopefully, from a unified point of view, there is much
more to be learned about computation.

Unity of these computational models is not surprising because physicists
generally believe that quantum field theory contains quantum mechanics, which
in turn contains classical
mechanics.

Correspondingly, we present the (quantum field computation) thesis:

\indent A. {\em Quantum field computation contains quantum computation as a
proper subset}.

\indent B. {\em Quantum field computation involves more powerful computational
tools than quantum computation (e.g. infinite dimensional Hilbert spaces, some
methods of real computation, and holomorphic functions)}.

On the other hand, quantum methods may not be suitable for certain real
computing problems.  
A simple
 argument is that mathematics is a
much wider subject than physics; not all mathematics is necessarily applicable
to physics.

At present, quantum field theory is regarded only as an asymptotically valid
theory~\cite{Weinberg}.
  We expect that well developed future generalizations
of physics theories, which include general relativity (Einstein gravitation),
and perhaps supersymmetry, will replace quantum field theory in the above
thesis, with additional and more powerful computational tools brought into
play.

\subsection{Quantum Computation}

Based on quantum theory, 
Feynman proposed {\it quantum computation}~\cite{Feynman}.  
  In
quantum computation, as opposed to Turing computation, qubits (quantum bits)
are used in place of classical bits.  A bit could be in one of two discrete
states, $0$ or $1$.  A qubit, on the other hand, corresponds to a $2$ level
quantum system, like a spin $1/2$ state of an electron.  We can have a complex
linear superposition of wave functions (eigenstates) of a $2$ level quantum
system so that two complex number amplitudes (in ${\mathbb C}^2$) are
involved in each qubit.  Feynman hoped to exploit the quantum system itself by
making it do the computation, but practically, decoherence {\it noise} is a
serious implementation problem even with very much less than $10$ qubits.

     As a result of Feynman's proposal 
there has been an enormous amount of research, not only on quantum computation,
but also on quantum cryptography.  The efforts in this regard are to seek
improved ways of performing computations, including a refinement
of Church's thesis by Deutsch~\cite{Deutsch}
 to tackle quantum computation, or
building new types of computing machines.  It is hoped that not only
exponentially faster computation will be achieved~\cite{Shor},
 but that better understanding of computational complexity will come about
~\cite{Vazirani, NielsenChuang}.

\subsection{Real Computation}

In another direction, the classical discrete digital Turing computer~\cite{tu}
 has been generalized to include the possibility of computing over
the continuum~\cite{sm2,BCSS}.
  This generalization is called {\it
real computation}.  The need for doing this is because 
computing over the continuum is more appropriate to the way we do analysis,
physics, numerical analysis and engineering problems.  Accordingly, the
classic logical theory of computation was enhanced with analysis, topology
and algebraic geometry.

Until recently, it was considered unthinkable to speak of computing over a
continuum, for example, over the infinite number of points in the real interval
$[0,1]$, without approximating at a finite number of points.  But Tarski,
in a little known paper~\cite{Tarski}, 
 proved completeness over the reals for
elementary algebra and geometry.  The complexity was extremely high
(exponential), but Smale {\it et al}~\cite{sm2,BCSS} have rectified that
situation.  Tarski's result is in contrast to G\"odel's famous
theorem~\cite{Papa} of incompleteness of arithmetic over
the integers ${\mathbb Z}$, and to Turing's theorem~\cite{tu} 
 of undecidability
of the Halting problem for computation over the integers ${\mathbb Z}$.

A question was raised by Penrose~\cite{Penrose}
 as to whether the Mandelbrot set
was an (albeit beautiful, picturesque) example of an undecidable set (i.e. a
recursively enumerable set that is not recursive).  It was concluded that it
was not possible to answer this question because there was no proper
definition for computing over the continuum.  One problem is: how does one
feed a real number, consisting of an infinitely long sequence of bits, into a
computing machine in finite time?  A proper definition was indeed given in the
work of Smale {\it et al}~\cite{sm2,BCSS} on real computation, and the
question on the Mandelbrot set was answered in the affirmative.  (The proof
hinges on the fact that the Hausdorff - Besicovitch dimension of the boundary
of the Mandelbrot set is indeed equal to $2$.)

In fact, Tarski hoped to build a machine which would compute over the reals. 
But it is now possible to do some simple real computation even on a Turing
machine.  Our thesis on quantum field computation relies heavily, not only on
the fact that quantum field theory generalizes quantum theory, but also on
possibilities of computing over the continuum.

Real computation is a computing model that is based on classical mechanics
and classical dynamical systems.  But classical mechanics could also be
extended to include relativity, resulting in relativistic
mechanics~\cite{Shapiro}. 

\subsection{Quantum Field Computation}

In studying atomic phenomena, classical mechanics has been
replaced by quantum mechanics.  Correspondingly the classical computer could
be improved with a quantum computer.  But we could also think of more general
models of computation 
based on adding relativity to quantum theory to get
relativistic quantum field theory, and consider appropriate quantum field
computation models.

In an approach to the central computer science problem of the P
(Polynomial time) versus NP (Nondeterministic Polynomial time)~\cite{Papa}) 
complexity classes, Freedman proposed a {\it quantum field
computer}~\cite{Freedman}.  Under consideration were topological quantum
field theories, and physical systems which contained non-Abelian gauge
terms in the Lagrangian.  The initial preparation of states was supposed to be
consistent with {\it knot} types~\footnote
 {The user-friendly,
interactive and animated color graphics ``SnapPea" program for creating knots
and studying hyperbolic 3-manifolds is available at:
http://thames.northnet.org/weeks/index/SnapPea.html .}.

Of course, in a general situation, as in non-Abelian gauge theories,
string theories (including general relativity),
superstring
theories, or topological field theories~\cite{Witten},
 quantum field
computation would be an immensely difficult undertaking. 

 But due
to the work of Wightman on relativistic quantum field theory (incorporating
Einstein's special relativity and employing analytic functions of several
complex variables)
 many of the components for some
quantum field computation are already available.
There is extensive literature on Wightman's model, for example,
in~\cite{sw, Jaffe, Kazhdan}
 as well as the references cited there.  Some additional 
computation methods are described in~\cite{ACMCombi}.  
  We concentrate therefore on explaining the unity among
computation models.

In this article the approach
is based on mathematical physics but the results also impact computer
science.

\section{Relationships Between Computation Models}

There is a 
remarkable relationship between quantum field computation and 
real computation.  Computation over the continuum appears in quantum field
computation as well as in real computation.  In the former, it is already
possible to compute over {\it cells} which are actually certain chunks
of the continuum space ${\mathbb C^n}$ of $n$ complex variables.

We might say this comes about because it is natural to consider a physical or
quantum system in the continuum limit.  In fact Isaac Newton,
 when studying gravitation, found it natural to consider a continuous
distribution of matter to model the earth's gravitational action at external
points.  From continuum quantum mechanics, by combining relativity, we have
quantum field theory, a system with an infinite number of degrees of freedom. 
The development in real computation of the Newton endomorphism method in
numerical analysis follows naturally from Newton's continuum limit.

Real computation could also be regarded as a stepwise
form of {\em analog computation\/} working within a continuum.

Conversely, quantum (mechanics) computation would be suspected to be a discrete
finite case of quantum field computation where the number of qubits is finite, 
and the corresponding Hilbert vector space is a finite dimensional vector
space.  

At the present time, quantum
computation proceeds as a time evolution over a finite number of discrete time
intervals, whereas time must be regarded as a continuous variable.  But
space and time are interwoven in relativity, depending on the frame of
reference: thus the need to handle the problem in a covariant manner.  Also, 
because a quantum field computation model does exist, it is important to say
that quantum computation can therefore benefit by including considerations
of relativity, methods of computing over the continuum, and an
unbounded number of qubits (infinite dimensional Hilbert space).

The important concepts for quantum computation are unitary transformations,
finite superposition of states, entanglement, and quantum cryptography. 
Superposition is standard also in quantum field theory.  Entanglement is a
rather interesting form of superposition, with applications to quantum
teleportation considerations and quantum cryptography, and relies on a basis
of EPR (named after Einstein, Podolsky and Rosen) or Bell states.	
The EPR
{\it gedanken} (thought) experiment itself, from the point of view of quantum
measurement theory, is not further discussed in quantum computation theory
because the absence of hidden variables is now an accepted fact.

Just as the rotation group is of importance in non-relativistic quantum
mechanics (with Euclidean geometry), the Lorentz group (with Minkowski
geometry) is relevant to relativistic quantum mechanics.  The Lorentz
group contains the rotation group as a subgroup.

Thus a basic symmetry
group in quantum mechanics is SU(2), the special (determinant $= 1$) unitary
group of $2\times 2$ complex matrices.  This is also the universal covering
group of the rotation group (real special orthogonal group) SO(3) in
3-dimensional space.
  
SU(2) is a proper subgroup of  SL(2,$\mathbb C$), the universal covering
group of the Lorentz group, which is the symmetry group for relativity in the
usual 1-time and 3-space dimensions.  Hence SL(2,$\mathbb C$) is the group
appropriate for quantum field computation.

Consider, for example, the EPR states. 
One particular EPR state, based on electron spins,	
can be written as
 $$\lbrace |01\rangle \mbox{}- |10\rangle \rbrace /\sqrt 2,$$
 or equivalently as
 $$\lbrace |\uparrow\downarrow\rangle \mbox{}-|\downarrow\uparrow\rangle
\rbrace /\sqrt 2,$$
where, in the usual description, the
first qubit refers to Alice and the second to Bob.
They prepared the entangled state, perhaps on Earth when
they were together, and now Bob could be in the Alpha Centauri system, at a
space-like separation from Alice on Earth.  This is simply the singlet state
for adding two spins of $1/2$, where we have a simultaneous eigenstate of the
total spin angular momentum $S=0$, and the total z-component of spin $S_z=0$.
(It is
possible also to have entangled states for photons, which can have horizontal
or vertical polarization.  Entanglement produces quantum interference between
photons.)

Addition of angular momentum of spins $1/2$ appears here as $D^{1/2} \times
D^{1/2} = D^1 + D^0$, in terms of decomposition of representations of the
rotation group in 3- dimensional space.  In quantum field computation, this
group is enlarged to the group SL(2,$\mathbb C$), which covers the Lorentz
group.  In general, one considers irreducible representations $D^{(j/2, k/2)}$
of SL(2,$\mathbb C$).  

The concept of electron spin $1/2$ is added on to (non-relativistic) quantum
computation in an {\em ad hoc} fashion.  But Dirac showed that electron spin
naturally follows from considerations of relativity, and the requirement of
first order differential equations~\cite{Bethe}.  
Entanglement of quantum
states is also applicable to the relativistic theory, i.e. to quantum field
theory.  Indeed, experiments verify that quantum rather than classical
field theory gives the correct results~\cite{Mandel, Zeilinger}.

\section{Field Theory Enhancements to Quantum Computation}

  Quantum field theory not only includes all of
quantum mechanics, and classical mechanics, but much more in
the form of well-known results.	 
Examples are, discrete
anti-unitary symmetries, CPT invariance, and the spin statistics
connection~\cite{sw}.  Our purpose
here is to exploit results that enhance quantum computation, through working
with a relativistic quantum field theory model.

Non-relativistic quantum mechanics is not complete because radiative
corrections have to be made to it, using field theory.  In dealing with a
system corresponding to an infinite number of degrees of freedom, it is well
known historically that formulations of quantum field theory like perturbation
theory lead to infinities resulting in the need for renormalization. 
Nevertheless, quantum electrodynamics has turned out to be ``the most
accurate theory known to man"~\footnote{This statement is attributed to
Feynman.}.  Dirac, Schwinger and Feynman are some of the
principal contributors to quantum electrodynamics
(the
spectacular history of which is related in~\cite{SSS})
 and hence to quantum field
theory~\cite{Weinberg}.
  Relativistic covariance is of paramount importance in
correctly performing the renormalization process.

If there is some way we can avoid approximations due to series
expansions in perturbation theory, and also avoid renormalization problems, at
least up to our point of departure of computational enhancements, we should do
so.  Fortunately
we can achieve this by working within the Wightman
formulation~\cite{sw, Jaffe, Kazhdan}
 of quantum field theory.  
Reasons for the
fruitfulness and utility of this formulation, from a current perspective, are
available in~\cite{Kazhdan}.
   We are dealing with fields in the Heisenberg
picture without using perturbation theory nor any particular time frame
related Hamiltonians.  The theory is in terms of analytic functions (Wightman
functions) of several complex variables. These functions arise from their
boundary values which are vacuum expectation values (in Dirac's {\it bra -
ket} notation)  of the form  $${\cal W}_m(x_1, x_2, \ldots x_m) = \\
  \langle\Omega | \phi_1(x_1) \phi_2(x_2) \ldots \phi_m(x_m) | \Omega\rangle$$
of products of $m$ quantum field operators in a
separable Hilbert space. The field operators transform according to
appropriate unitary spin representations of the Poincar\'e (inhomogeneous
$SL(2,{\mathbb C})$) group, for $3+1$ space-time dimensions and generally
transform as spinors in $s$-dimensions.	 
Spinor indices
have been suppressed here, but are available in~\cite{sw}.  Wightman 
reconstructs quantum fields uniquely from these analytic functions.  This is
called the {\it reconstruction theorem}. 

Let the ($m$-point) Wightman function be denoted by $W(n; z)$ where $z$
denotes the set of $n$ complex variables. Here, $n = sm$ where $s \geq 2$ is
the space-time dimension; space-time will consist of 1-time and $(s-1)$-space
dimensions.  
  $m$ is also called the function order, and $n$ will be called the
function index.
  It was only recently known, how to
physically understand concepts like closed time-like loops in more than one
time dimension~\cite{Gog},
 where the second time dimension is in a tiny
loop of a Kaluza - Klein type brane universe theory.  However, we will
restrict ourselves here to the conventional single dimension in
time~\cite{Hawking}.
  We use a general space-time dimension $s$ for the sake
of considering {\it uniformity} of computation (to approach {\it universality}
of computation), for what appear to be computational problems in their own
right; whereas certain values of $s$, such as $2,3,4,5,10,11$ and $26$ have
turned out to be more appropriate for purely physics problems.

Because these analytic functions are fundamental to the theory, one is led to
computation of holomorphy domains for these functions over the space of
several complex variables, ${\mathbb C}^n$.  
  The mass spectrum is
assumed to be reasonable in the sense that momentum vectors $p^{\mu}$ lie in
the closed forward light cone, with time component $p^0 > 0$ except for the
unique vacuum state having $p = 0$.

Thus the many complex numbers (or amplitudes)
that need to be handled in quantum field computation were indeed tamed as
complex variables in analytic functions, i.e. the Wightman functions. 
(However, these are
not the same entanglement type complex amplitudes used in quantum
computation.)

Computation over ${\mathbb C}^n$ is common also in real computation.  But, in
the Wightman model, we seem to have stronger computation because of the use
of holomorphic functions (over ${\mathbb C}^n$) of several complex
variables.  Not only the physics of quantum theory and special relativity, but
also microcausality is utilized.

\section{Analog Computation and Symbolic Computation}
It is interesting that the strengths of analog and symbolic computation come
into play as quantum field computation supplements and enhances quantum
computation.  We think that, it is sometimes debatable as to what is symbolic
or analog computation when it comes to computing over the continuum.  In real
computation there seems to be a subtle re-emergence of the old analog
computer in a new and powerful form.  This new form is effectively digitally
clamped to avoid noise problems (such as voltage drifts in potentiometers)
which plagued the old analog computer.

When $s=2$, i.e. in 1-dimensional space and 1-dimensional time, 
deterministic exact analog computation~\cite{ACMCombi}
(computation over 
{\it cells} in the continuum of ${\mathbb C}^n$) is used to obtain what are
called  primitive extended tube domains of holomorphy for $W(n; z)$.
The computation can be done with 
essentially reversible logic, as a Horn clause satisfiability problem
(HORNSAT),
and simulating on a Turing machine.  But HORNSAT is in the complexity class P
(polynomial time)~\cite{Papa}.  This is now a deterministic problem of
complexity P, but also implies non-deterministic polynomial time computation,
in the complexity class NP, as discussed below.

We note a couple of points in this connection.
First, we rely here on the soundness theorem and the converse theorem,
namely, G\"odel's theorem of completeness of first order predicate
calculus~\cite{Papa}.  Secondly, reversibility of computation is an asset
because  information content is maximized, or equivalently, the entropy
increase is minimized.

Just as the classical computer, {\it Turing machine}, computes over ${\mathbb
Z}$ or (up to polynomial time) equivalently over ${\mathbb Z}_2$ (the
classical bit representation of numbers), we now have what can be called a
{\it complex Turing machine}, in fact, a {\it severally complex Turing
machine}. 

The primitive extended tube domains are bounded by analytic hypersurfaces,
namely several Riemann cuts,
and other analytic hypersurfaces of types denoted by $S$ and $F$, which too
play a role.
  These domains are in
the form of {\it semi-algebraic sets} in the language used in real
computation.  Since the computation is symbolic, it is also exact, which is
important in handling holomorphic functions.  

Because of Lorentz invariance properties of the physics involved, the domains
have a structure referred to as {\it Lorentz complex projective
spaces}.
  (These Lorentz complex spaces are different, but
physical, ``non-Euclidean" views of complex projective spaces which are
well known in mathematics.) 
  Related to this invariance are certain
continuum {\it cells} over which the computation occurs.  Thus this
computation is also like analog computation which would otherwise be regarded
as impossible to do exactly.

In this simple case, it is possible to think that (suitably encoded) pieces or
whole continuous group orbits are being fed into the Turing machine.  Hopefully
there will be more possibilities like this in the future.

\subsection{Analytic Extensions}

In relativistic quantum field theory it is possible to implement the physical
requirement of microcausality.  There exists quantum microcausality (field
operators commute or anti-commute) at space-like separations.

Together with the consequence of permutation invariance of the domains, the
edge-of-the-wedge theorem provides enlargements of the original
primitive domains of analyticity into analyticity in unions of permuted
primitive domains.

Mapping these union domains creates some Boolean satisfiability
problems. In fact, the novel methods of computation raise interesting issues of
computability and complexity.
 
\subsection{Non-deterministic Holomorphic Extensions}

By the nature of analytic domains in more than one complex variable, it is in
general possible to further extend these domains towards the maximal domains
called {\it envelopes of holomorphy}.  By considering boundary related
semi-algebraic sets, there are non-deterministic computations of holomorphic
extensions of domains.  After the guessing step, the verification is by 
deterministic processes mentioned above.  Historically, this method was used
by K\"all\'en and Wightman in computation, for the first time, of the
holomorphy envelope for $m = 3$.  

Because HORNSAT is in P, ~\cite{Papa},
this results in an NP type problem, i.e.
guessing the result and verifying in polynomial time.
  So this part of the problem is in the
complexity class NP.  (We note also that HORNSAT is P-complete.)

Built-in permutation invariance 
has considerable power 
just as $n!$ rapidly dominates over $2^n$ for large $n$.  In applying local
commutativity, it might appear that we have to generate permutations of $m$
objects; in fact, no algorithm is known to do this in polynomial
time.  But because of the power of non-deterministic computation~\cite{Papa},
we are allowed to guess a candidate for a permutation; and then we can verify,
in polynomial time, whether the guess is indeed a permutation, throwing out the
candidate in case it is not a permutation.

\section{Uniformity of Computation}

Uniformity in the direction of universal computation has been
discussed~\cite{BCSS}, in different contexts, including numerical analysis.
 We do have certain types of uniformity here.

First we note that the computation is independent of any particular form
of Lagrangian or dynamics, and is uniform in $n$, qualifying for a universal
quantum machine over ${\mathbb C}^\infty$.  The latter space is defined in 
Appendix B.

\subsection{Function Index Uniformity}
When the logic program runs for $s=2$, dynamic memory allocation is used
through the operating system.  Because $n$ can be input as a variable, only
part of the whole memory management cost is outside the program.  The program
itself is independent of $n = sm$ and therefore is uniform in $n$, which is
unbounded above.  We can call this {\it function index uniformity in}
$n^\infty$. 

\subsection{Space-time Dimension Uniformity}

In addition, there is uniformity in the dimension $s \geq 2$ of space-time, in
the following manner.
Given a dimension $s \ge 2$ of space-time, looking at the
semi-algebraic sets defining the primitive extended tube domains of
holomorphy (with hypersurface boundaries) and at function orders, there are
three different classes of orders.  These classes comprise, a) lower order W
functions,  b) intermediate order W functions, and c) high order W
functions.
  Extended tube domains for all high order W
functions have the same complicacy.  For a)
we have $m \le s + 1$, and for c), $m > s(s-1)/2 + 2$.  The remaining
cases lie in class b).  For example, there is no class b) for $s = 2$ (i.e.
class b) is empty), the most complicated primitive domain being for the 3-point
function.  If $s = 3$, then $m = 5$ is the only case in class b).  When $s =
4$, we have in class b), the cases, $m = 6, 7$ and $8$.

Since $s \ge 2$ is unbounded above, we can call this {\it space-time dimension
uniformity in}
 $s^\infty$.

\subsection{Uniformity of WHOLO}

We recall that although deterministic complexity classes are closed under
complements, the non-deterministic complexity class NP is not necessarily
closed under complements. In fact, it is known~\cite{Papa} that the
complexity class P is a subset of both complexity classes co-NP and NP.  Also
the problem PRIMES (``given an integer, is it a prime?") belongs to both
complexity classes co-NP and NP.  But it is not known whether PRIMES belongs
to the complexity class P i.e. no polynomial time algorithm is known for
PRIMES.  This lack of knowledge is the basis for the success of trapdoor
cipher type encryption algorithms like RSA.

Let us denote the Wightman problem of computing holomorphy envelopes with the
notation WHOLO.  Thus we have seen above that the problem WHOLO has
uniformity in $n^\infty$ and $s^\infty$. 

The holomorphy envelopes
for different orders $m$ of Wightman
functions are related; 
the holomorphy envelope for order $m$ is contained
in the intersection of holomorphy envelopes
for lower order functions~\cite{ACMCombi}.
(This is
further explained in Appendix A.)

For example, in $s=2$, the 4-point function cannot be continued beyond the
2-point function Riemann cuts nor the (permuted) 3-point function
K\"all\'en-Wightman domains of holomorphy.

This is a statement regarding analyticity that does not exist, and thus refers
to the complements of domains of holomorphy;  hence the use of the prefix
{\it co-}.  Because computations of analytic extensions of domains are
non-deterministic (hence the notation {\em N}), we can say that we have 
{\it
co-N} uniformity over $s^\infty$, and in particular, co-NP
complexity for $s=2$.  

In the case that the holomorphy domains are Schlicht, which is the only case
known at present in this quantum field model, then 
the domains of holomorphy in Appendix A are closed under complements.
This implies, in $s=2$ for the relevant part of the WHOLO problem, that the
succinct certificates (or polynomial witnesses) of co-NP complexity for higher
order functions are contained in those for lower order functions.  This could
have implications regarding problems which are in co-NP and not in NP.

\section{Discussion}
We have not used the non-linear positive definiteness conditions for W-
functions in Hilbert space.  These conditions are required for the
reconstruction theorem.  On the other hand, we want to exploit the complexity
conditions for the linear-program problem as computational problems in their
own right.

The original problem posed by Freedman~\cite{Freedman} for a quantum field
computer, was motivated by the existence of a great deal of mathematical
physics relating to the case $s=3$.  In this 3-dimensional space-time, space
itself is 2-dimensional, and there are a host of fruitful statistical
mechanics and field theory problems in this case~\cite{Jaffe}.  For
example, instead of particles having to be Bosons or Fermions as in $s=4$, we
have {\em Anyons} corresponding to braid-group statistics.  (The knot problem
and 3-dimensional manifolds studied as knot complements, show up here.)  There
is also the fractional quantum Hall effect, which not only has produced some of
the most accurate experimental results to date, but is the fertile testing
ground for new physical theories as well.  In particular, Chern-Simons type
gauge interaction terms in the Lagrangian~\cite{Deser} give more insight into
field theories, including gravitation.  In the future, we should expect such
theories to be part of quantum field computation.

At the time of Turing, a {\it computer} was a human being doing calculations. 
In the present era, {\it computers} are machines on which humans are extremely
dependent, not only for calculations but also for modeling natural phenomena. 
Quantum computers indeed have the potential of greater power than classical
computers.  Exploiting real computation methods and quantum field computation
enhancements by invoking special relativity, gives even stronger computational
tools.  In quantum cryptography, more powerful computation means stronger
private code distribution
and weaker public code methods.
In the private code case, when Eve eavesdrops on the
transmission of quantum information from Alice to Bob, the quantum data is
disturbed so that Bob can decide it is so and discard those data items,
requesting Alice to re-transmit.
In the public code case, for
example in the well-known RSA encryption and decoding algorithm, the code
will be easier to break. 

There is discrete translational invariance in quantum computation,
compared to continuous translational invariance in quantum field
computation.  The discrete Fourier transform is of profound importance to
the power of quantum computation.  In the early days of quantum field
theory, it was usual to quantize over a finite, rather than an infinite, box. 
The finite box incorporates discrete translational invariance and allows
discrete Fourier transforms.

Since, in quantum field theory, particles with arbitrary spins can be
annihilated and created, we can talk about {\it qubits, qutrits, ququads,
...}, and in general, about {\it quspinors}. 

Relying on a fruitful set of models, we have related what appeared to be
different models of quantum and classical computation based on relativistic
and non-relativistic quantum mechanics and classical mechanics.  Exact
deterministic and non-deterministic computation over continuous domains appear
naturally.  Furthermore there is uniformity in computation over, unbounded
above, or arbitrarily high index $n$ of $W(n;z)$ and arbitrarily high
dimension $s$ of space-time.  

It is good to break up a complex problem into several parts and analyze the
complexity of each part separately.  Three parts of the problem WHOLO have been
identified above.  (There is a fourth part, namely, the representation
of unions of domains, which has been possible to do only by human
interaction.)  In the case $s=2$ the first part is in the complexity class P
(and is P-complete), the second in NP, and the third in co-NP.  

Identification, within quantum field computation, of these 
methods of computation raise interesting issues of computability and
complexity, and possibly could shed more light,
not only on computability, but also  
on the description of Nature by fundamental physics theories themselves.

\section{Conclusion}
By {\em unity} between computation models we mean that the models are actually
parts of a whole, higher (or broader) model of computation.  Viewed from such a
broader perspective it should be possible to better understand how the
different parts, namely different computation models, fit together.  The
situation here is quite analogous to the situation in physics theories, where
quantum field theory is the higher model (in this article), which contains
quantum mechanics.  Correspondingly we have quantum field computation as the
higher level model which contains quantum computation.

Although some parts of the Wightman model of quantum field theory are
exploited here, and in fact the only way employed up to the present of
connecting up with the real computational model, these parts of the Wightman
model should not be regarded as the only possible way of thinking in the
future.  The higher level model in physics is now quantum field theory, but
this model might need to be expanded later (by including more symmetry 
groups, general relativity, topological fields, etc).

Each mathematical physics theory could possibly have some interesting, 
novel, computational and complexity ramifications.  This idea was suggested
by Freedman~\cite{Freedman}.  Accordingly, within quantum field theory we have
identified P versus NP consequences and certain uniformities of computation. 
These uniformities are helpful in thinking of universality of computation, a
hopeful problem for the future.

Through Einstein's relativity, we have shown why there is unity between
quantum field computation, real computation (computation over the continuum)
and quantum computation.  The Church Turing thesis for computation is
supposed to be presently enhanced with the quantum field computation thesis we
have proposed above.  Thus the ingeneous methods in quantum
computation, of dealing with discrete Fourier transforms, entangled states and
fault-tolerant quantum error corrections could be profitably supplemented
with concepts of infinite dimensional Hilbert spaces and (some) methods of
computation over the continuum.

\appendix
\section{Relations Between Holomorphy Envelopes}

The holomorphy envelopes $H[D_m]$ for different orders $m$ of Wightman
functions are related
 in the following way.

{\it For} $0 < r < m$, and {\it relative to} ${ H[D_m] }$,
$${ H[D_m] \subset \\ 
 \bigcap_{\sigma \in {\rm P}_m} \\  
  \left\{ H [\sigma D_{m-r}] \times 
  (\sigma {\mathbb C}^{sr})  \right\}  },$$ 

where $\sigma$ denotes permutations in ${\rm P}_m$, the permutation group in
the $m$ points of the $m$-point W function.  
This is a theorem which is referred to in~\cite{ACMCombi}.
In the case of {\it Schlicht}
domains (analogous to single sheeted Riemann surfaces in $\mathbb C$), the
$\subset$ sign means set theoretic inclusion (but more subtle otherwise).

\section{The complex space ${\mathbb C}^\infty$.}
This space is defined to be the infinite disjoint union
$\bigsqcup^\infty_{m = 1}{\mathbb C}^{sm}$.  We have followed the type of
definition used in real computation~\cite{BCSS}.   There is a certain lattice
pattern in the values of $n$ allowed in the subsequence $n=sm$.  For example,
since $s \geq 2, m \geq 2$, n is never a prime number.  Of course ${\mathbb
C}^\infty$ involves holomorphy in the variables, which is a much stronger
condition than differentiability, so that the space is different from
$C^\infty$, the space of infinitely differentiable functions.

\end{document}